\documentclass[useAMS,usenatbib]{mn2e}

\usepackage{graphicx}
\usepackage{txfonts}

\def\D{\ifmmode\mathcal{D}\else$\mathcal{D}$\fi}

\title[Hierarchy and sizes of star formation regions]
      {Hierarchy and size distribution function of star formation 
       regions in the spiral galaxy NGC~628}

\author[A.~S.~Gusev]
       {Alexander~S.~Gusev \\
        Sternberg Astronomical Institute, Lomonosov Moscow State University,
        Universitetsky pr. 13, 119992 Moscow, Russia, 
        {\sf gusev@sai.msu.ru} \\
             }

\date{Accepted 2014 June 2. Received 2014 June 2; in original 
form 2013 December 24}

\begin{document}

\maketitle

\begin{abstract}
Hierarchical structures and size distribution of star formation 
regions in the nearby spiral galaxy NGC~628 are studied over a range of 
scale from 50 to 1000~pc using optical images obtained with 1.5~m telescope 
of the Maidanak Observatory. We found hierarchically structured 
concentrations of star formation regions in the galaxy, smaller regions with 
a higher surface brightness are located inside larger complexes having a 
lower surface brightness. We illustrate this hierarchy by dendrogram, or 
structure tree of the detected star formation regions, which demonstrates 
that most of these regions are combined into larger structures over several 
levels. We found three characteristic sizes of young star groups: 
$\approx65$~pc (OB associations), $\approx240$~pc (stellar aggregates) 
and $\approx600$~pc (star complexes). The cumulative size distribution 
function of star formation regions is found to be a power law with a slope of 
approximately $-1.5$ on scales appropriate to diameters of 
associations, aggregates and complexes. This slope is close to the slope 
which was found earlier by B.~Elmegreen~et~al. for star formation regions 
in the galaxy on scales from 2 to 100~pc.
\end{abstract}


\section{Introduction}

As is known, such physical processes as gravitational collapse and turbulence 
compression play a key role in creation and evolution of star formation 
regions over the wide range of scales, from star complexes over 
OB associations down to compact embedded clusters and to clumps of young 
stars inside them. These stellar systems form a continuous hierarchy 
of structures for all these scales \citep{efremov1995,efremov1998,
elmegreen2000,elmegreen2002,elmegreen2006b,elmegreen2011}. It is suggested 
that the hierarchy extends up to 1~kpc 
\citep*{efremov1987,elmegreen2006c,zhang2001}.

\citet{efremov1987} and \citet{ivanov1991} described at least three 
categories of hierarchical star groups on the largest levels: 
OB associations with a length scale $\approx80$~pc, stellar aggregates with 
a length scale $\approx250$~pc and star complexes with diameters 
$\approx600$~pc. H\,{\sc i}/H$_2$ superclouds are ancestors of star 
complexes; OB associations are formed from giant molecular clouds 
\citep{efremov1989,efremov1995,efremov1998,elmegreen1994,elmegreen2006c,
elmegreen2009,odekon2008,marcos2009}. Sizes and clustering of these 
structures have been studied for many nearby spiral and irregular galaxies 
\citep*{bastian2005,bianchi2012,battinelli1991,battinelli1996,borissova2004,
bresolin1996,bresolin1998,bruevich2011,elmegreen2001,feitzinger1984,
gouliermis2010,gusev2002,harris1999,magnier1993,pietrzynski2001,
pietrzynski2005,sanchez2010,wilson1991,wilson1992}. Power-law power spectra 
of optical light in galaxies suggest the same maximum scale, possibly 
including the ambient galactic Jeans length 
\citep*{elmegreen2003a,elmegreen2003b}. If the ambient Jeans length is the 
largest scale, then a combination of gravitational and turbulent 
fragmentations can drive the whole process. Observed star formation rates 
in galaxies can follow from such turbulent structures \citep{krumholz2005}.

Hierarchical clustering disappears with age as stars mix. The densest 
regions have the shortest mixing times and lose their substructures first. 
Nevertheless, very young clusters have a similar pattern of subclustering, 
suggesting that this structure continues down to individual stars 
\citep*{brandeker2003,dahm2005,heydari2001,nanda2004,oey2005,sanchez2013}.

\begin{figure*}
\resizebox{0.98\hsize}{!}{\includegraphics[angle=000]{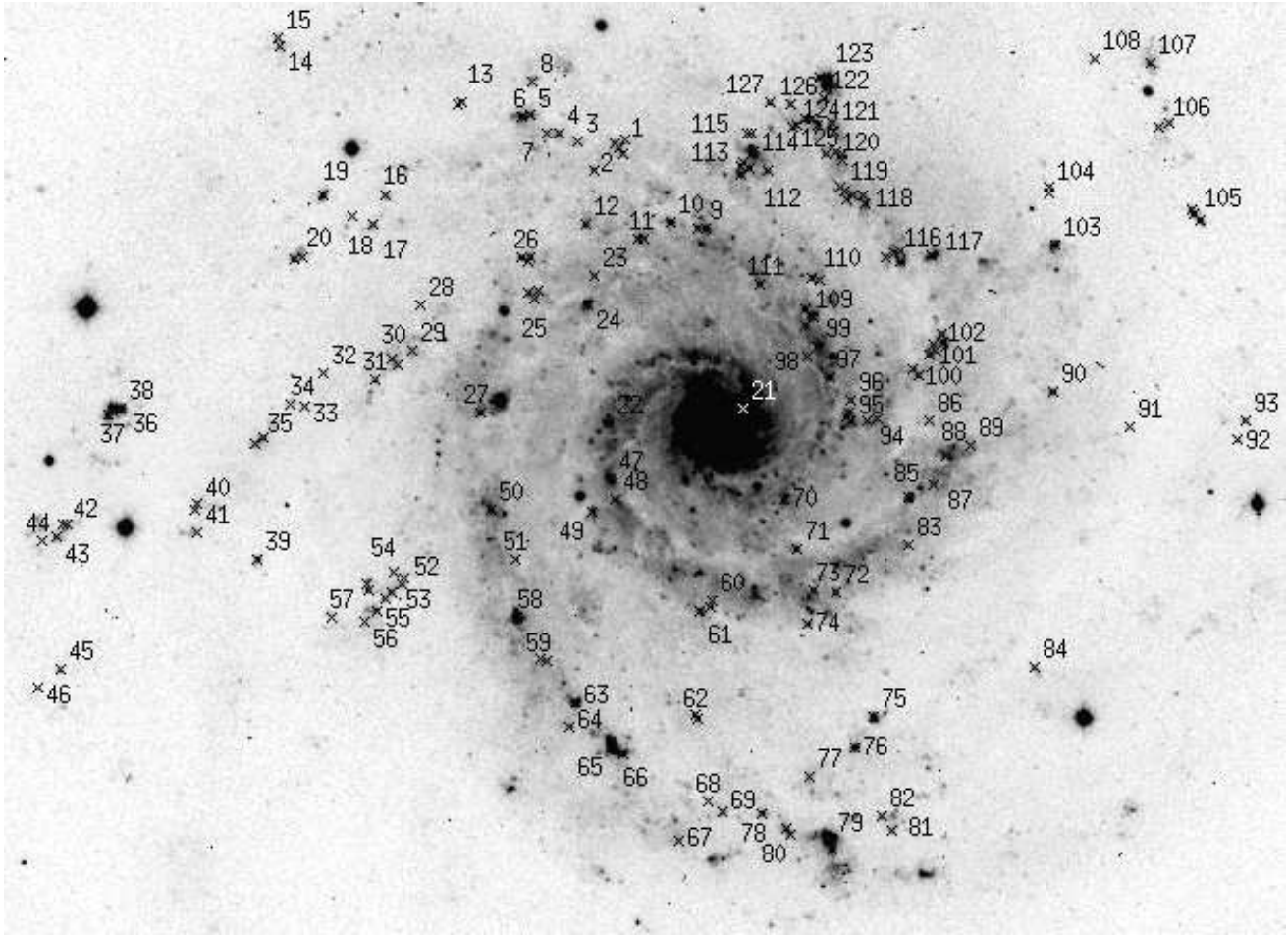}}
\caption{$B$ image of NGC~628 and positions of the galaxy's star formation 
regions (crosses). The numbers of the star formation regions from 
Table~\ref{table:positions} are indicated. The image size is 
$8.26\times6.00$~arcmin. North is upward and east is to the left.
}
\label{figure:fig_iden}
\end{figure*}

\begin{table}
\caption[]{\label{table:param}
Basic parameters of NGC~628.
}
\begin{center}
\begin{tabular}{ll} \hline \hline
Parameter                                & Value \\
\hline
Type                                     & Sc \\
RA (J2000.0)                             & 01$^h$36$^m$41.81$^s$ \\
DEC (J2000.0)                            & +15$\degr$47$\arcmin$00.3$\arcsec$ \\
Total apparent $B$ magnitude ($B_t$)     & 9.70 mag \\
Absolute $B$ magnitude ($M_B$)$^a$       & -20.72 mag \\
Inclination ($i$)                        & $7\degr$ \\
Position angle (PA)                      & $25\degr$ \\
Apparent corrected radius ($R_{25}$)$^b$ & 5.23 arcmin \\
Apparent corrected radius ($R_{25}$)$^b$ & 10.96 kpc \\
Distance ($D$)                           & 7.2 Mpc \\
\hline
\end{tabular}\\
\end{center}
\begin{flushleft}
$^a$ Absolute magnitude of a galaxy corrected for Galactic extinction and
inclination effect. \\
$^b$ Isophotal radius (25 mag\,arcsec$^{-2}$ in the $B$-band) corrected for 
Galactic extinction and absorption due to the inclination of NGC~628.
\end{flushleft}
\end{table}

The interstellar matter also shows a hierarchical structure from the largest 
giant molecular clouds down to individual clumps and cores. The complex 
hierarchical structure of the interstellar matter is shaped by supersonic 
turbulence \citep{ballesteros2007}. The scaling relations observed in 
molecular clouds \citep{larson1981} can be explained by the effect of 
turbulence, where energy is injected at largest scales and cascades down to 
the smallest scales, creating eddies and leading to a hierarchical structure 
on all scales \citep{elmegreen2006}. It is believed that turbulence plays a 
major role in star formation; it creates density enhancements that become 
gravitationally unstable and collapse to form stars \citep{elmegreen2006}. 
The spatial distribution of young stars and stellar groups on wide length 
scales probably reflects this process.

The purpose of this paper is to study size distribution and hierarchical 
structures of star formation regions in nearby face-on spiral 
galaxy NGC~628 (Fig.~\ref{figure:fig_iden}), based on our 
own observations in the $U$, $B$, and $V$ passbands. This 
galaxy is an excellent example of a galaxy with numerous star formation 
regions observed at different length scales. We use the term 
'star formation regions', which includes young star complexes, 
OB associations, H\,{\sc ii} regions, i.e. all young stellar groups 
regardless of their sizes.

\citet{hodge1976} identified 730 H\,{\sc ii} regions in the galaxy. 
\citet{ivanov1992} estimated sizes and magnitudes of 147 young stellar 
associations and aggregates in NGC~628 and discussed briefly hierarchical 
structures at the scales from 50 to 800~pc. \citet{larsen1999} studied 38 
young star clusters with an effective diameters from 2 to 90~pc. 
\citet{bruevich2007} obtained magnitudes, colours and sizes of 186 star 
formation regions based on the list of H\,{\sc ii} regions from 
\citet{belley1992}.

\citet{elmegreen2006} studied distributions of size and luminosity of 
star formation regions over a range of scales from 2 to 110~pc using 
progressively blurred versions of blue optical and H$\alpha$ images from 
the {\it Hubble Space Telescope (HST)}. They counted and measured 
features in each blurred image using SExtractor program and found that the 
cumulative size distribution satisfies a power law with a slope of 
approximately from --1.8 to --1.5 on all studied scales.

\begin{figure*}
\vspace{5.0mm}
\resizebox{0.92\hsize}{!}{\includegraphics[angle=000]{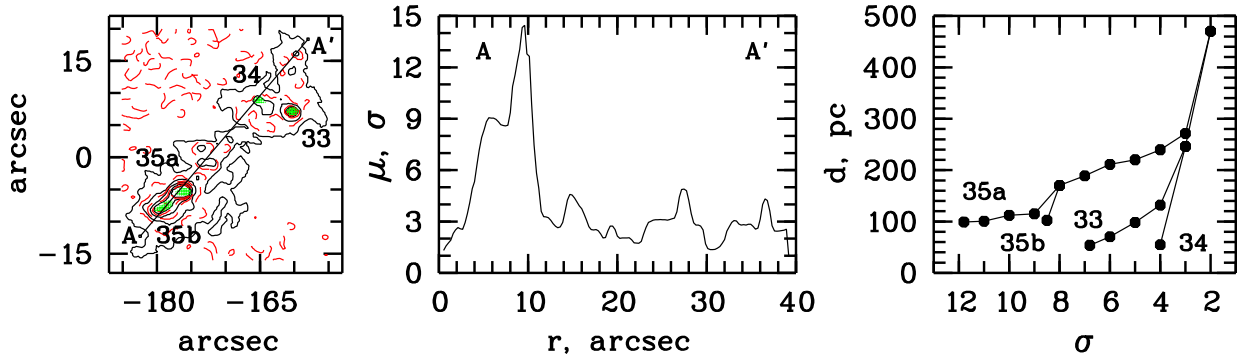}}
\caption{Left panel: contour map of the vicinity of star formation regions 
Nos.~33-35. 
Grey areas correspond to the regions Nos.~33, 34, 35a, and 35b within their 
half-maximum brightness level. Red dashed contour levels correspond to the 
levels of $\sigma$, $3\sigma$, $5\sigma$, $7\sigma$, and $9\sigma$, black 
solid contour levels correspond to the levels of $2\sigma$, $4\sigma$, 
$6\sigma$, $8\sigma$, and $10\sigma$ above the average brightness level of 
background. Position of profile A--A' is shown. Central panel: photometric 
profile A--A'. Surface brightness, $\mu$, is given in units of $\sigma$. 
Right panel: diameters of star formation regions Nos.~33-35 and their 
hierarchical structures 
measured at the different levels of surface brightness in units of $\sigma$.
}
\label{figure:fig33_35}
\end{figure*}

The fundamental parameters of NGC~628 are presented in 
Table~\ref{table:param}. We take the distance to NGC~628, obtained in 
\citet*{sharina1996} and \citet*{vandyk2006}. We used the position angle and 
the inclination of the galactic disc, derived by \citet{sakhibov2004}. 
Other parameters were taken from the LEDA data 
base\footnote{http://leda.univ-lyon1.fr/} \citep{paturel2003}. We adopt the 
Hubble constant $H_0 = 75$ km\,s$^{-1}$Mpc$^{-1}$ in the paper. With the 
assumed distance to NGC~628, we estimate a linear scale of 
34.9~pc\,arcsec$^{-1}$.

Observations and reduction stages of $UBVRI$ images for NGC~628 have 
already been published in \citet{bruevich2007}. The reduction of the 
photometric data was carried out using standard techniques, with the European 
Southern Observatory Munich Image Data Analysis 
System\footnote{http://www.eso.org/sci/software/esomidas/} ({\sc eso-midas}).

\section{Identification and size estimations of star formation regions}

In \citet{bruevich2007}, we have identified star formation regions in 
the galaxy with the list of H\,{\sc ii} regions of \citet{belley1992}, 
based on their H$\alpha$ spectrophotometric data. The list of 
\citet{belley1992} is still the most complete survey of H\,{\sc ii} regions 
and their parameters in NGC~628. Note that our coordinate grid coincides 
with that of \citet{kennicutt1980} and is systematically shifted with respect 
to that of \citet{belley1992}. Altogether, we identified 127 of 132 star 
formation regions studied in \citet{belley1992}. Three regions 
\citep[Nos. 1, 2, and 96 in][]{belley1992} were outside the field of view of 
our images. Two star formation regions (Nos. 23 and 76) are missing in the 
list of \citet{belley1992}. \citet{belley1992} did not distinguish between 
isolated star formation regions, with typical sizes about 60-70~pc, and 
compound multi-component regions, with typical sizes about 200~pc. We obtained 
images of the galaxy with better seeing than \citet{belley1992}. As a result, 
we were able to resolve the compound star formation regions into components.

Firstly, we identified such subcomponents by eye. We selected the 
components, the maximal (central) brightness in which was at least 3 times 
higher than the brightness of surrounding background. Next, we fitted 
profiles of star formation regions using Gaussians. The components separation 
condition was 
that the full width at half-maximum (FWHM) of the region is less than the 
distance between centres of Gaussians. Numbers of these complexes in the 
first column of Table~\ref{table:positions} contain additional letters: 
'a', 'b', 'c', and 'd'. Compound regions which do not satisfy this 
condition were classified as objects with observed, but unresolved, internal 
structure. In total, we identified 186 objects (Fig.~\ref{figure:fig_iden}).

In this paper we use the numbering order adopted in \citet{bruevich2007}. 
It coincides with the numbering order of \citet{belley1992} with the 
exception of the missed star formation regions.

We found that 146 regions from Table~\ref{table:positions} have a star-like 
profile (see the last column in this table). Other 40 objects have a 
non-star-like (extended (diffuse) or multi-component) profile, i.e. these 
objects have an observed, but unresolved, internal structure.

We took the geometric mean of major and minor axes of a star formation 
region for the star formation region's characteristic diameter $d$:
$d = \sqrt{d_{max} \times d_{min}}$. We measured $d_{max}$ and $d_{min}$
from the radial $V$ profiles as the FWHM for regions having a star-like 
profile, or as the distance between points of maximum flux gradient for 
regions having non-star-like profiles. We adopted seeing for the uncertainty 
in the size measurements, which definitely exceeds all other errors. Obtained 
parameters of star formation regions are presented in 
Table~\ref{table:positions}.

\section{Hierarchical structures of star formation regions}

The simplest way to study hierarchical clustering is to identify 
structures of different hierarchical levels based on lower level surface 
brightness thresholds above the background level. The similar method 
was used by \citet{gouliermis2010}, who used the stellar density levels 
to study hierarchical stellar structures in the dwarf irregular galaxy 
NGC~6822. They identified hierarchical structures using density thresholds 
$1\sigma - 5\sigma$ above the average background density level with step of 
$1\sigma$.

However, this direct way is not applicable for identification of hierarchical 
structures in NGC~628. The background level varies significantly in the 
galactic plane. The surface brightness of the background differs by several 
times inside spiral arms and in interarm regions.

Therefore we modified the technique of \citet{gouliermis2010}. 
Identification and size estimation of 186 star formation regions at the 
highest hierarchical level (Level~1) were done using their half-maximum 
brightness levels, independent of background levels (see Section~2). 
Additionally, we fitted the profiles of star formation regions along their 
minor and major axes using Gaussians. To identify structures of Level~2 and 
lower, we measured the background surface brightness in the $V$ passband in 
the vicinity of every group of star formation regions of Level~1.

\begin{figure}
\vspace{3.1mm}
\resizebox{0.90\hsize}{!}{\includegraphics[angle=000]{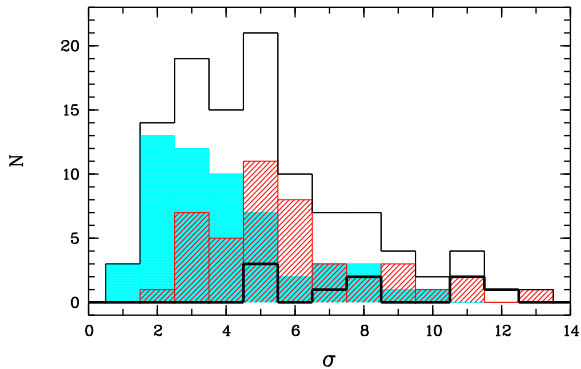}}
\caption{Distribution histogram of star formation regions of Levels~2-5 by 
the level of maximum brightness decrease. Brightness is given in units of 
$\sigma$. Grey histogram is the distribution of star formation regions of 
the lowest hierarchical level. Shaded histogram is the distribution of 
star formation regions of the first hierarchical level from the lowest one. 
Thick black histogram is the distribution of star formation regions of the 
second hierarchical level from the lowest one. See the text for details.
}
\label{figure:fig_sigma}
\end{figure}

The selection of a threshold in units of $\sigma$ above the average 
brightness level of background for star formation regions of Level~2 was 
carried out based on two basic conditions: (i) it must be lower than the 
level of brightness of the appropriate star formation region of Level~1 and 
(ii) it must deviate more than 4 pixels (seeing of the $V$ image) from the 
fitting Gaussian of the profile of the star formation region at Level~1. The 
same conditions were applied to select the brightness level of every next 
lower level of the hierarchy. The exception was made for several resolved 
close binary star formation regions, such as 40a-40b, where the second 
condition is not applied. To identify star formation regions of lower 
hierarchical levels, we used lower levels of brightness.

To select surface brightness thresholds, we firstly analysed a 
typical light distribution in selected star formation regions and their 
vicinities. Example of such region, star formation regions Nos.~33-35, is 
given in Fig.~\ref{figure:fig33_35}.

Fig.~\ref{figure:fig33_35} (central panel) shows that the surface brightness 
falls irregularly with distance from the knots of star formation: 
'plateau-like' areas with constant surface brightness alternate with 
areas of a sharp drop in brightness. At such sites, a fall in brightness 
usually exceeds $1\sigma$ value. At higher hierarchical levels, where 
the surface brightness is higher, absolute drop in brightness is larger than 
at lower levels of the hierarchy. As a result, diameters of star formation 
regions increase slowly with a decrease of brightness level within the same 
hierarchical level. Significant growth of the diameters is observed only at 
merger of two separate star formation regions into one common star formation 
region at the lower hierarchical level (Fig.~\ref{figure:fig33_35}).

We consider brightness in units of $\sigma$. So the brightness level, where 
the maximum brightness decrease is observed, is also measured in units of 
$\sigma$. Maximum brightness decrease corresponds to the minimum of 
first derivative of the brightness profile function 
(Fig.~\ref{figure:fig33_35}, central panel) in units of $\sigma$. After 
measuring the brightness level in units of $\sigma$ by the maximum 
brightness decrease, we determine size of star formation region with the 
isophots as described in Section~2.

We analysed all hierarchical structures in vicinities of star formation 
regions of Level~1 and determined which level of brightness corresponds to 
the level of maximum brightness decrease in them 
(Fig.~\ref{figure:fig_sigma}).

Distribution of star formation regions of Levels~2-5 by the level of maximum 
brightness decrease shows two maxima at $3\sigma$ and $5\sigma$ 
(Fig.~\ref{figure:fig_sigma}). Distribution of star formation regions of the 
lowest hierarchical level has a maximum at $2\sigma-3\sigma$, distribution of 
star formation regions 
of the first hierarchical level from the lowest one shows maxima at 
$3\sigma$ and $5\sigma-6\sigma$. Star formation regions of the second 
hierarchical level 
from the lowest one have characteristic levels of maximum brightness decrease 
of $5\sigma$, $8\sigma$, and $11\sigma$ (Fig.~\ref{figure:fig_sigma}).

Analysis of the distribution of star formation regions by the level of 
maximum brightness 
decrease, in units of $\sigma$, has shown that neither arithmetic nor 
geometric sequences of the brightness levels are suitable to describe the 
hierarchical structures of star formation regions. When using a geometric 
sequence, we may 
miss some of the hierarchical levels. When using an arithmetic sequence, we 
lose some of the brightness levels because they do not satisfy the condition 
(ii) (Fig.~\ref{figure:fig33_35}). In this case, low hierarchical levels 
will correspond to arbitrary levels of brightness.

Analysis of the distribution showed that the best sequence of brightness 
levels is the Fibonacci sequence, $1\sigma$, $2\sigma$, $3\sigma$, 
$5\sigma$, $8\sigma$, as an intermediate sequence between arithmetic and 
geometric sequences.

Diameters of star formation regions of the lower hierarchical levels which 
have the maximum 
brightness decrease at the level of $4\sigma$ or $6\sigma-7\sigma$ are measured 
at the next lower surface brightness level of $3\sigma$ or $5\sigma$, 
respectively. Typically, the difference between diameters measured at the 
levels of $3\sigma$ and $4\sigma$, or $5\sigma$ and $6\sigma-7\sigma$ does 
not exceed 35-40~pc, a value of the seeing of the image 
(Fig.~\ref{figure:fig33_35}).

Thus, we used surface brightness thresholds of $8\sigma$, $5\sigma$, 
$3\sigma$ and $2\sigma$ above the average brightness level of background 
in the vicinity of star formation region. The threshold of $1\sigma$ was not 
used due to large fluctuations of background around many identified groups 
of star formation regions. 

For each individual region, not every next lower brightness level 
satisfies the conditions adopted. Such brightness levels were missed. 
Furthermore, a full set of brightness levels from $8\sigma$ to $2\sigma$ 
above the background was used only for star formation regions Nos.~79a and 
79b and hierarchical 
structures of a lower order related with them (Table~\ref{table:tree}). The 
lowest level of every hierarchical structure usually corresponds to the 
brightness level of $2\sigma$ or $3\sigma$ above the background 
(Fig.~\ref{figure:fig_sigma}). As a result, the same hierarchical level may 
correspond to different levels of brightness.

Diameters of star formation regions at Levels~2 and lower were measured in 
the same manner as for star formation regions of Level~1: 
$d = \sqrt{d_{max} \times d_{min}}$, where $d_{min}$ and $d_{max}$ are 
diameters along the major and minor axes of star formation region.

\begin{figure*}
\resizebox{0.98\hsize}{!}{\includegraphics[angle=000]{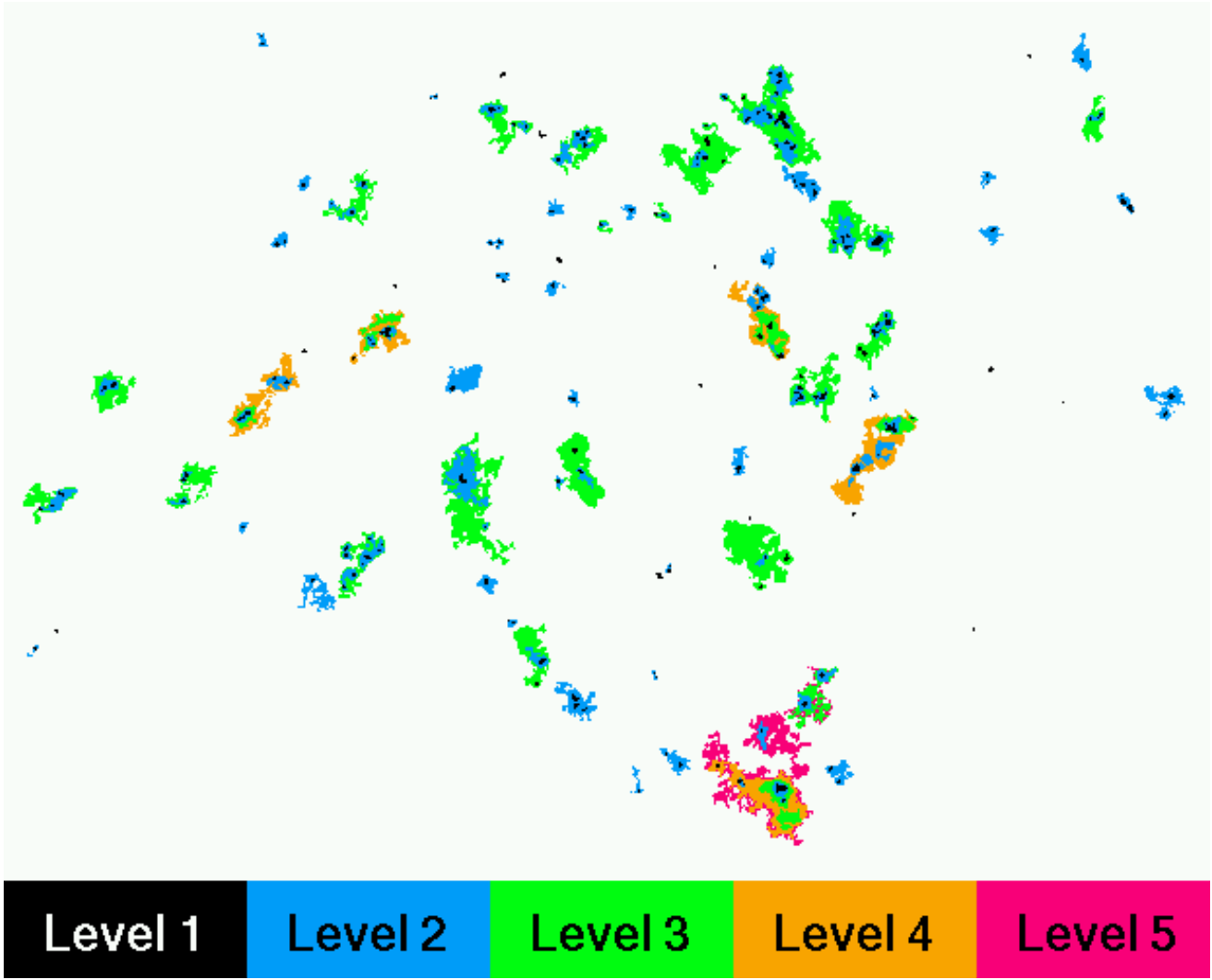}}
\caption{Map of star formation regions of different levels of the hierarchy. 
Regions of higher levels of the hierarchy are shaded darker than lower 
ones. The image size is $8.26\times6.00$~arcmin. North is upward and east is 
to the left.
}
\label{figure:fig_levels}
\end{figure*}

Star formation regions obtained on different hierarchical levels, 
and their sizes are presented in Table~\ref{table:tree}. Some star 
formation regions of low hierarchical levels consist of one or several 
star-like cores (star formation regions of Level~1) and an extended halo. 
Such star formation regions are indicated by letter 'h' in 
Table~\ref{table:tree}. A map of location of these objects in 
the galactic plane is shown in Fig.~\ref{figure:fig_levels}.

To illustrate the hierarchical structures we used so-called dendrograms.
Dendrograms were introduced as 'structure trees' for analysis of 
molecular cloud structures by \citet{houlahan1992}, refined by 
\citet{rosolowsky2008}, and used in \citet{gouliermis2010} to study 
hierarchical stellar structures in the nearby dwarf galaxy NGC~6822. A 
dendrogram is constructed by cutting an image at different brightnesses and 
identifying connected areas, while keeping track of the connection to 
surface brighter smaller structures (on a higher level) and surface fainter 
larger structures (on the next lower level, which combines structures of the 
previous level). 

\begin{figure}
\vspace{2.3mm}
\resizebox{1.00\hsize}{!}{\includegraphics[angle=000]{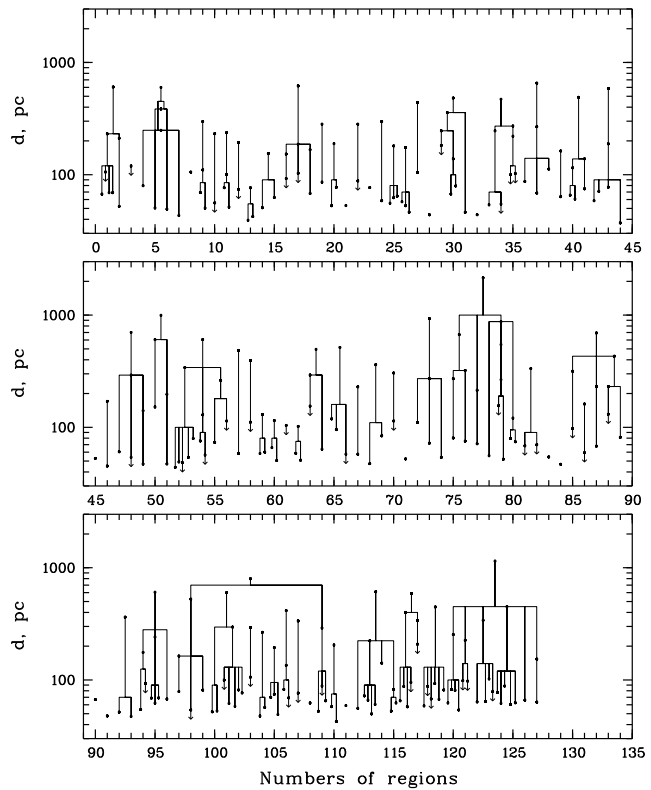}}
\caption{Dendrogram of star formation regions structures. The black dots 
indicate star formation regions from Tables~\ref{table:positions}, 
\ref{table:tree}. Regions which are united into a hierarchical structure 
are connected by solid line. The numbering order might not strictly follow 
the order of hierarchical structures (see the text for details). The arrows 
down indicate star formation regions with an observed internal structure 
(star formation regions with a non-star-like profile).
}
\label{figure:fig_tree}
\end{figure}

The dendrogram for the star formation regions from 
Tables~\ref{table:positions} and 
\ref{table:tree} is presented in Fig.~\ref{figure:fig_tree}. Unlike 
\citet{gouliermis2010}, we constructed the dendrogram using the ordinate 
axis in units of diameter. It better illustrates length scales of 
hierarchical structures. The combination of this dendrogram with the map of 
Fig.~\ref{figure:fig_levels} illustrates graphically the hierarchical spatial 
distribution of star formation regions in NGC~628.

The dendrogram demonstrates that most star formation regions are combined 
into larger 
structures over, at least, 1-2 levels. We found only 12 separate associations 
without visible internal structure, which are out of hierarchical structures 
(Fig.~\ref{figure:fig_tree}). Most of them are located in interarm regions 
(Fig.~\ref{figure:fig_iden}). The largest ($d>1$~kpc) and the most populous 
(8-17 star formation regions of Level~1) structures are located in the ends 
of spiral arms. 
First of them (Nos.~75-80) is located near the corotation radius, which was 
obtained in \citet{sakhibov2004} based on a Fourier analysis of the spatial 
distribution of radial velocities of the gas in the disc of NGC~628. Largest 
and brightest in UV star complex of the galaxy was found here in 
\citet*{gusev2014}. Second structure (Nos.~120-127) is located in the 
nothern-western part of NGC~628, in the disturbed part of the spiral arm 
(Fig.~\ref{figure:fig_iden}).

As seen from the dendrogram, the numbering order does not reflect 
correctly the hierarchical structures. The numbering is violated for star 
formation regions 
Nos.~4-7 at Level~2, Nos.~85-89 at Level~4, and Nos.~97-109 at Level~4 
(Table~\ref{table:tree}).

\section{Size distributions of star formation regions}

\begin{figure}
\vspace{3.1mm}
\resizebox{1.00\hsize}{!}{\includegraphics[angle=000]{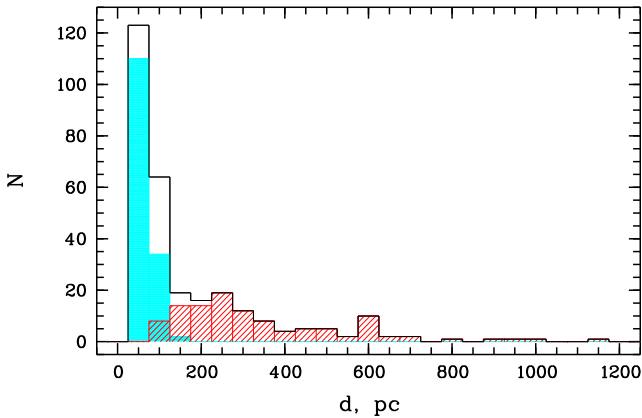}}
\caption{Number distribution histograms of all star formation regions 
from Tables~\ref{table:positions} and \ref{table:tree}, star formation 
regions of Level~1 with a star-like profile (grey histogram), and 
star formation regions of Levels~2 and lower (shaded histogram).
}
\label{figure:fig_hist}
\end{figure}

In Fig.~\ref{figure:fig_hist}, we present size distribution histograms for 
three sets of star formation regions under study. The first set includes 297 
regions of all hierarchical levels, the second set is a sample of 146 
associations with a star-like profile, and the third set includes 111 regions 
of Level~2 and lower from Table~\ref{table:tree}. The second set unites the 
star formation regions without an observed internal structure; their 
subcomponents (if exist) have sizes $\le35-40$~pc. The third set includes 
only star formation regions with obvious internal structure; their 
subcomponents were detected and measured.

As seen from the figure, associations with a star-like profile have a 
narrow range of sizes, from 40 to 100~pc, with a few exceptions. The mean 
diameter of these star formation regions is equal to $66\pm18$~pc. This is a 
typical size of OB associations. Star formation regions with extended profile 
have, on average, slightly larger sizes, $\sim100$~pc. As a result, the size 
distribution of star formation regions of Level~1 with both star-like and 
extended profiles is displaced a little toward the larger sizes (see 
Fig.~\ref{figure:fig_hist} and Table~\ref{table:mean}).

Star formation regions of lower levels clearly show a bimodal size 
distribution. Two maxima 
at $\approx250$ and $\approx600$~pc are observed (Fig.~\ref{figure:fig_hist}). 
The first smoothed peak corresponds to a characteristic size of stellar 
aggregates by classification of \citet{efremov1987}, and the second 
peak is located on diameters, which are typical for star complexes.

\begin{table}
\caption[]{\label{table:mean}
Diameters of star formation regions.
}
\begin{center}
\begin{tabular}{ccc} \hline \hline
Star formation   & $d^a$      & $d^b$ \\
regions          & (pc)       & (pc)  \\
\hline
Associations$^c$ &  $66\pm18$ &  64 \\
Associations$^d$ &  $72\pm26$ &  66 \\
Aggregates       & $240\pm90$ & 234 \\
Complexes        & $583\pm84$ & 601 \\
\hline
\end{tabular}\\
\end{center}
\begin{flushleft}
$^a$ Mean diameter. $^b$ Diameter obtained from best fitting Gaussian. \\ 
$^c$ Associations with a star-like profile (146 objects). \\ 
$^d$ All associations from Table~\ref{table:positions} (186 objects).
\end{flushleft}
\end{table}

We also fitted size distributions of studied sets of star formation regions 
using Gaussians. To fit the size distribution for the set of 111 complex 
star formation regions, we used a combination of two Gaussians. It was found 
that all sets of star formation regions have size distributions close to the 
Gaussian distribution. Diameters obtained from the best-fit Gaussians are 
almost the same as the mean diameters for all sets of star formation regions 
(Table~\ref{table:mean}).

\begin{figure}
\vspace{3.1mm}
\resizebox{1.00\hsize}{!}{\includegraphics[angle=000]{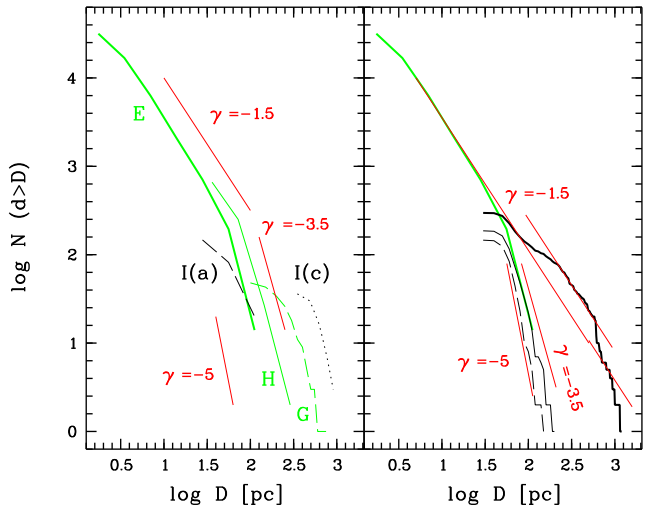}}
\caption{Left panel: cumulative size distribution functions for regions of 
\citet{elmegreen2006} (E; grey thick solid curve), H\,{\sc ii} regions of 
\citet{hodge1976} (H; grey thin solid curve), associations (I(a); black 
dashed curve) and complexes (I(c); black dotted curve) of \citet{ivanov1992}, 
large star formation regions in the spiral arms of the galaxy (G; grey thin 
dashed curve) from \citet{gusev2014}. Right panel: cumulative size 
distribution functions for regions of \citet{elmegreen2006} (grey thick solid 
curve), 146 star formation regions with a star-like profile from 
Table~\ref{table:positions} 
(black dashed curve), 186 star formation regions from 
Table~\ref{table:positions} (black solid curve), and 297 star formation 
regions from Tables~\ref{table:positions} and \ref{table:tree} 
(black thick curve). Dark thin solid straight lines in both panels represent 
slopes $\gamma$~=~--1.5, --3.5 and --5 of the size distribution function. See 
the text for details.
}
\label{figure:fig_fsize}
\end{figure}

Following \citet{elmegreen2006}, we constrained the cumulative size 
distribution function in the form $N (d>D) \propto D^{\gamma}$, where $N$ is 
the integrated number of objects that have a diameter $d$ greater than some 
diameter $D$ (Fig.~\ref{figure:fig_fsize}). 

Detailed exploration of the size distribution of objects in NGC~628 was 
made in \citet{elmegreen2006} in the range of scales from 2 to 
110~pc\footnote{For an adopted distance of 7.2~Mpc.} based on {\it HST} 
images. For regions in the central part of the galaxy brighter than the 
$3\sigma$ noise limits in $B$ and $V$ images, \citet{elmegreen2006} found 
that the cumulative size distribution obeys a power law, with a slope 
$\gamma \approx -1.5$ in the range from 2 to 55~pc. The similar slope of the 
cumulative size distribution function was found for OB associations from the 
list of \citet{ivanov1992} in the range from 30 to 110~pc. The size 
distribution of larger objects, H\,{\sc ii} regions studied by 
\citet{hodge1976}, satisfies a power law with a slope $\gamma \approx -3.5$ 
in the range from 100 to 300~pc. The size distribution of large 
star formation regions (in the range from 300 to 600~pc) in 
spiral arms of NGC~628 obtained in \citet{gusev2014} shows a slope 
$\gamma \approx -4.5$. The size distribution 
of complexes from \citet{ivanov1992} gives $\gamma = -4.1$ in the range from 
500 to 1000~pc (Fig.~\ref{figure:fig_fsize}).

Summarizing the results of the size distribution obtained previously, we can 
conclude that the size distributions of star formation regions with a 
diameter of $\le100$~pc satisfy a power law with $\gamma \approx -1.5$. The 
distribution of larger star formation regions obeys a power law with 
$\gamma$ between $\approx -5$ and $\approx -3.5$.

In Fig.~\ref{figure:fig_fsize} (right panel) we present size distribution 
functions constructed for three sets of star formation regions. The first set 
includes 297 regions of all hierarchical levels, the second set is a sample 
of 186 star formation regions of Level~1, and the third set includes 146 
regions of Level~1 with star-like profile.

Size distribution of 146 star formation regions with a star-like profile, 
beginning with 
$d \approx 50$~pc, obeys a power law with a slope $\gamma \approx -5$. Size 
distribution of all 186 star formation regions of Level~1 satisfies a power 
law with a slope 
$-4\le\gamma\le-3.5$ in the range from 50 to 170~pc. It repeats the 
distribution of H\,{\sc ii} regions of \citet{hodge1976} with a displacement 
$\log D \approx 0.2$ (Fig.~\ref{figure:fig_fsize}). In general, the size 
distribution of star formation regions of Level~1 has slopes between 
--5 and --3.5, such as size distributions of previously studied star 
formation regions of a single level of hierarchy 
(Fig.~\ref{figure:fig_fsize}).

Note, that the end of the size distribution curve for regions of 
\citet{elmegreen2006} coincides with the beginning of the size distribution 
curve for our 186 star formation regions of Level~1 
(Fig.~\ref{figure:fig_fsize}). Given 
that the area studied in \citet{elmegreen2006} occupies $\sim 70\%$ of the 
area of NGC~628, which is studied in this paper, we can conclude that (i) the 
number of H\,{\sc ii} regions identified in \citet{belley1992} is smaller 
than the numbers of regions found by \citet{elmegreen2006} using SExtractor, 
and, that is more important, (ii) our measurements of sizes of star 
formation regions using photometric profiles are in a good agreement with 
measurements of \citet{elmegreen2006}.

More interesting behaviour is observed for the curve of size distribution of 
star formation regions of all hierarchical levels. It continues the size 
distribution curve 
for regions of \citet{elmegreen2006} at $d=30-40$~pc and has the same slope 
$\approx -1.5$ in the range from 45 to 85~pc -- diameters of OB associations. 
Flatter slope, $\gamma > -1$, is observed in the range from $\approx 90$ to 
$\approx 180$~pc for regions of Level~1 with an extended profile and for 
the smallest regions of Level~2. Size distribution of star formation regions, 
which are 
classified as stellar aggregates and complexes, obeys a power law with 
$\gamma = -1.5$ very well (see the distribution curve in the range from 190 
to 600~pc in Fig.~\ref{figure:fig_fsize}). Largest hierarchical structures 
with $d=0.65-0.9$~kpc are also distributed by sizes by a power law with 
$\gamma \sim -1.5$ (Fig.~\ref{figure:fig_fsize}).

Thus, the size distribution of star formation regions of all hierarchical 
levels continues the size distribution function for regions of 
\citet{elmegreen2006} towards the larger sizes with the same slope 
$\approx -1.5$.

\section{Discussion}

The modern theory of star formation explains an existence of OB associations 
and star complexes, which are associated by unity of an origin with hydrogen 
superclouds and giant molecular clouds, respectively 
\citep{elmegreen2006c,efremov1998}. Structures of H$_2$ on the intermediate 
scale length are unknown. However, such intermediate young stellar structures 
are observed in galaxies. These are stellar aggregates with diameters 
$\sim200-300$~pc.

Our Fig.~\ref{figure:fig_hist} shows a bimodal size distribution of star 
formation regions of 
Level~2 and lower. Bimodal size distributions with a secondary peak at 
$d=150-300$~pc were found for 'associations' in SMC \citep{battinelli1991}, 
M31 \citep{magnier1993}, NGC~2090, NGC~2541, NGC~3351, NGC~3621 and NGC~4548 
\citep{bresolin1998}, NGC~1058 and UGC~12732 \citep{battinelli2000}, NGC~300 
\citep{pietrzynski2001}, NGC~3507 and NGC~4394 \citep{vicari2002}, NGC~7793 
\citep{pietrzynski2005}. \citet{pietrzynski2005} named such 'associations' 
'superassociations'.

Thus, the existence of stellar structures, 'aggregates' or 
'superassociations', with a characteristic size 200--300~pc is confirmed by 
numerous obsevations in different galaxies. However, the question of origin 
of stellar aggregates is still open.

As we noted above, the size distribution of star formation regions of all 
hierarchical levels 
continues the size distribution of regions of \citet{elmegreen2006} with the 
same slope $\approx -1.5$ for sizes from 45~pc to $\sim0.9$~kpc. However, the 
function of size distribution deviates from a power law with the slope --1.5 
at $d=90-180$~pc and $600-650$~pc (Fig.~\ref{figure:fig_fsize}).

We believe that the flatter slope in the range of 90 to 180~pc is a result 
of significant number of star formation regions with a diameter of 
$\sim100-150$~pc with an 
unresolved internal structure. Taking into account such undetected objects will shift 
the distribution curve upward along the ordinate axis at sizes smaller than 
or equal to diameters of these star formation regions.

Opposite situation is observed at $d=600-650$~pc. Largest structures with 
$d>600$~pc have a low boundary surface brightness. They are difficult for 
identification in spiral arms of grand design galaxy NGC~628 because of the 
significant variations of background level (see Section~4). Underestimation 
a number of star formation regions at lowest hierarchical levels leads to a 
drastic drop in the size distribution curve.

In spite of small statistics, largest star formation regions with 
$d\approx0.65-0.9$~kpc are 
also distributed by size by a power law with $\gamma \sim -1.5$. It can be an 
additional argument in favor of the assumption of \citet{efremov1987}, 
\citet{elmegreen2006c} and \citet{zhang2001}, who adopted that the 
hierarchical structures extend to the scale of 1~kpc.

Taking into account the hierarchy of star formation regions is crucial for 
construction of the cumulative size distribution function. Neglecting the 
internal structure of star formation regions of higher hierarchical levels 
and underestimation of the number of star formation regions of lower 
hierarchical levels leads to a decrease or an increase of the slope of the 
size distribution function, respectively. To illustrate this, we compare 
size distributions of regions of \citet{elmegreen2006}, our star formation 
regions of all hierarchical levels, and our star formation regions of Level~1 
with any profiles in the range of scale from 50 to 110~pc in 
Fig.~\ref{figure:fig_fsize}.

On the scale of 200--600~pc, the characteristic sizes of stellar 
aggregates and complexes, the size distribution function has a constant 
slope. We believe that the sample of objects at different levels of 
hierarchy within this range of scale is complete.

The slope $\gamma$ of the cumulative size distribution function for star 
formation regions is of 
fundamental importance. It is associated with the fractal dimension of objects 
in the galaxy at different scales. \citet{elmegreen2006} introduced the 
fractal of dimension $\D\ $, where $\D\ = - \gamma$. Following 
\citet{elmegreen2006}, we believe that the size distribution of stellar groups 
suggests a fractal distribution of stellar positions projected on the disc 
of the galaxy, with a constant fractal dimension of $\D\ \approx 1.5$ in the 
wide range of length scales from 2~pc to 1~kpc. It is comparable to the 
fractal dimension of projected local interstellar clouds, 
$\D\ \approx 1.3$ \citep*{falgarone1991}, and to the fractal dimension of 
H\,{\sc i} ($\D\ = 1.2-1.5$) in the M81 group of galaxies 
\citep{westpfahl1999}.

\section{Conclusions}

We studied hierarchical structures and the size distribution of
star formation regions in the spiral galaxy NGC~628 over a range of 
scale from 50 to 1000~pc based on size estimations of 297 star formation 
regions. Most star 
formation regions are combined into larger structures over 
several levels. We found three characteristic sizes of young star groups: 
OB associations with mean diameter $d=66\pm18$~pc, stellar aggregates 
($d=240\pm90$~pc) and star complexes ($583\pm84$~pc).

The cumulative size distribution function of star formation regions satisfies 
a power law with a slope of $-1.5$ at scales from 45 to 85~pc, from 190 
to 600~pc, and from 650~pc to 900~pc, which are appropriate to the sizes of 
associations, aggregates and complexes. Together with the result of 
\citet{elmegreen2006}, who found the slope $-1.8\le\gamma\le-1.5$ for 
regions at scales from 2 to 100~pc, our result shows that the size 
distribution of young stellar structures in the galaxy obeys a power law with 
a constant slope of $\approx-1.5$ at all studied scales from 
$\approx2$~pc to $\approx1$~kpc.

Ignoring the hierarchical structures, i.e. using star formation regions of 
only one of hierarchical levels to examine the size distribution, gives 
slopes $-5\le\gamma\le-3$.

\section*{Acknowledgements}

The author is grateful to the referee for his/her constructive comments. 
The author is grateful to Yu.~N.~Efremov (Sternberg Astronomical Institute) 
for useful discussions. The author thanks E.~V.~Shimanovskaya (Sternberg 
Astronomical Institute) for help with editing this paper. The author 
acknowledges the usage of the HyperLeda 
data base (http://leda.univ-lyon1.fr). This study was supported in part by 
the Russian Foundation for Basic Research (project nos. 12--02--00827 and 
14--02--01274).

\appendix

\section{Parameters and hierarchical structures of star formation regions}

\begin{table*}
\caption[]{\label{table:positions}
Identification, offsets, and diameters of star formation regions.
}
\begin{center}
\scriptsize{
\begin{tabular}{cccccccccccccccccccc} \hline \hline
ID & ID       & N--S$^b$ & E--W$^b$ & $d$  & Note & & 
ID & ID       & N--S     & E--W     & $d$  & Note & & 
ID & ID       & N--S     & E--W     & $d$  & Note \\
   & (BR)$^a$ & (arcsec) & (arcsec) & (pc) &      & & 
   & (BR)     & (arcsec) & (arcsec) & (pc) &      & & 
   & (BR)     & (arcsec) & (arcsec) & (pc) &      \\
\cline{1-6}\cline{8-13}\cline{15-20}
  1a &   3 &  +108.8 &  --40.3 &  65 & st$^c$ & &
 48  &  51 &  --29.1 &  --40.1 &  55 &        & & 
 97  & 102 &   +17.8 &   +42.9 &  80 & st     \\
  1b &   3 &  +106.9 &  --38.5 & 105 &        & & 
 49  &  52 &  --33.6 &  --48.9 &  45 & st     & &
 98  & 103 &   +25.8 &   +33.8 &  55 &        \\
  1c &   3 &  +104.2 &  --37.4 &  70 & st     & &
 50  &  53 &  --32.8 &  --88.3 & 150 & st     & &
 99  & 104 &   +30.4 &   +38.3 &  80 & st     \\
  1d &   3 &  +110.4 &  --36.6 &  70 & st     & &
 51  &  54 &  --52.3 &  --78.5 &  45 & st     & &
100a & 105 &   +21.0 &   +75.1 &  50 & st     \\
  2  &   4 &   +98.4 &  --48.6 &  50 & st     & &
 52a &  55 &  --57.1 & --126.2 &  45 & st     & &
100b & 105 &   +18.9 &   +77.0 &  55 & st     \\
  3  &   5 &  +109.0 &  --54.5 & 120 &        & &
 52b &  55 &  --62.2 & --122.7 &  50 & st     & &
101a & 106 &   +27.2 &   +81.0 & 100 &        \\
  4  &   6 &  +112.2 &  --61.9 &  80 & st     & &
 52c &  55 &  --59.5 & --121.7 &  50 &        & &
101b & 106 &   +28.2 &   +83.4 &  60 & st     \\
  5  &   7 &  +119.7 &  --72.9 &  50 & st     & &
 53a &  56 &  --67.5 & --129.4 &  55 & st     & &
102a & 107 &   +30.6 &   +82.9 &  60 & st     \\
  6  &   8 &  +119.2 &  --76.1 &  50 & st     & &
 53b &  56 &  --64.8 & --127.0 &  80 & st     & &
102b & 107 &   +31.7 &   +86.9 &  80 & st     \\
  7  &   9 &  +112.8 &  --67.0 &  45 & st     & &
 54a &  57 &  --64.3 & --135.8 &  75 & st     & &
102c & 107 &   +34.9 &   +86.1 &  75 & st     \\
  8  &  10 &  +132.8 &  --72.3 & 105 & st     & &
 54b &  57 &  --61.6 & --136.3 &  55 &        & &
103  & 108 &   +69.0 &  +129.8 & 105 &        \\
  9a &  11 &   +75.5 &   --4.9 &  70 & st     & &
 55  &  58 &  --72.0 & --132.6 &  75 & st     & &
104a & 109 &   +89.6 &  +127.7 &  50 & st     \\
  9b &  11 &   +76.0 &   --8.6 &  50 & st     & &
 56  &  59 &  --76.6 & --136.9 & 115 &        & &
104b & 109 &   +92.0 &  +127.7 &  55 & st     \\
 10  &  12 &   +78.1 &  --18.7 &  55 &        & &
 57  &  60 &  --74.4 & --149.9 &  60 & st     & &
105a & 110 &   +78.6 &  +186.1 &  70 & st     \\
 11a &  13 &   +71.7 &  --31.3 &  75 & st     & &
 58  &  61 &  --74.4 &  --77.9 & 110 &        & &
105b & 110 &   +81.1 &  +183.4 &  75 & st     \\
 11b &  13 &   +72.0 &  --28.9 &  50 & st     & &
 59a &  62 &  --91.0 &  --68.9 &  60 & st     & &
105c & 110 &   +83.2 &  +182.9 &  50 & st     \\
 12  &  14 &   +77.1 &  --51.5 &  75 &        & &
 59b &  62 &  --91.5 &  --67.0 &  60 & st     & &
106a & 111 &  +115.2 &  +170.3 &  80 & st     \\
 13a &  15 &  +124.0 & --101.1 &  40 & st     & &
 60a &  63 &  --70.7 &   --3.3 &  65 & st     & &
106b & 111 &  +116.8 &  +174.3 &  70 &        \\
 13b &  15 &  +124.2 &  --99.3 &  40 & st     & &
 60b &  63 &  --68.6 &   --2.7 &  50 & st     & &
107  & 112 &  +139.4 &  +166.6 &  75 &        \\
 14  &  16 &  +145.8 & --170.2 &  50 & st     & &
 61  &  64 &  --72.6 &   --7.3 & 105 &        & &
108  & 113 &  +141.0 &  +145.0 &  60 & st     \\
 15  &  17 &  +149.3 & --170.7 &  65 & st     & &
 62a &  65 & --112.3 &   --9.4 &  60 & st     & &
109a & 114 &   +37.8 &   +33.5 &  55 & st     \\
 16  &  18 &   +88.3 & --128.9 &  90 &        & &
 62b &  65 & --114.2 &   --8.6 &  50 & st     & &
109b & 114 &   +41.6 &   +36.2 &  90 &        \\
 17  &  19 &   +77.3 & --133.7 & 105 &        & &
 63  &  66 & --107.8 &  --55.5 & 155 &        & &
109c & 114 &   +44.2 &   +33.0 &  65 & st     \\
 18  &  20 &   +80.3 & --141.9 &  70 & st     & &
 64  &  67 & --116.8 &  --57.7 &  65 & st     & &
110a & 115 &   +56.8 &   +35.9 &  60 & st     \\
 19  &  21 &   +88.8 & --153.1 &  85 & st     & &
 65a &  68 & --122.4 &  --42.2 & 120 & st     & &
110b & 115 &   +56.0 &   +38.6 &  45 & st     \\
 20a &  22 &   +63.7 & --164.6 &  55 & st     & &
 65b &  68 & --125.4 &  --41.7 &  95 & st     & &
111  & 116 &   +54.4 &   +15.7 &  60 & st     \\
 20b &  22 &   +64.8 & --161.4 &  75 & st     & &
 66  &  69 & --127.2 &  --37.4 &  60 &        & &
112  & 117 &   +98.1 &   +18.6 &  55 & st     \\
 21  &  24 &    +5.8 &    +9.5 &  55 & st     & &
 67  &  70 & --160.8 &  --15.8 &  60 & st     & &
113a & 118 &   +96.5 &    +8.2 &  70 & st     \\
 22  &  25 &    +0.2 &  --43.0 &  90 &        & &
 68  &  71 & --145.6 &   --4.6 &  45 & st     & &
113b & 118 &   +99.7 &    +8.2 &  65 & st     \\
 23  &  26 &   +57.6 &  --48.6 &  75 & st     & &
 69  &  72 & --149.9 &    +1.3 &  85 & st     & &
113c & 118 &  +101.3 &    +9.3 &  50 & st     \\
 24  &  27 &   +46.1 &  --51.0 &  60 & st     & & 
 70  &  73 &  --28.8 &   +25.3 & 115 &        & &
113d & 118 &   +99.2 &   +11.4 &  60 & st     \\
 25a &  28 &   +50.9 &  --73.7 &  55 & st     & &
 71  &  74 &  --48.3 &   +29.8 &  50 & st     & &
114  & 119 &  +105.0 &   +13.0 & 140 & st     \\
 25b &  28 &   +51.4 &  --69.9 &  60 & st     & &
 72  &  75 &  --65.4 &   +45.0 & 110 & st     & &
115a & 120 &  +112.2 &   +11.1 &  55 & st     \\
 25c &  28 &   +48.8 &  --71.3 &  65 & st     & &
 73  &  77 &  --64.3 &   +36.5 &  70 & st     & &
115b & 120 &  +112.2 &   +12.5 &  65 & st     \\
 26a &  29 &   +64.2 &  --76.6 &  55 & st     & &
 74  &  78 &  --77.1 &   +34.3 &  55 & st     & &
116a & 121 &   +64.8 &   +64.5 &  65 & st     \\
 26b &  29 &   +64.8 &  --73.4 &  55 & st     & &
 75  &  79 & --113.4 &   +59.7 &  80 & st     & &
116b & 121 &   +66.1 &   +67.5 &  90 & st     \\
 26c &  29 &   +62.6 &  --73.7 &  45 & st     & &
 76  &  80 & --125.1 &   +52.5 &  75 & st     & &
116c & 121 &   +66.9 &   +70.3 &  60 & st     \\
 27  &  30 &    +4.5 &  --92.3 & 105 & st     & &
 77  &  81 & --136.6 &   +34.6 &  70 & st     & &
116d & 121 &   +62.6 &   +70.3 &  95 &        \\
 28  &  31 &   +46.4 & --115.8 &  45 & st     & &
 78  &  82 & --150.4 &   +16.7 &  55 & st     & &
117  & 122 &   +65.3 &   +82.3 & 210 &        \\
 29  &  32 &   +28.5 & --118.5 & 180 &        & &
 79a &  83 & --160.0 &   +42.6 & 155 &        & &
118a & 123 &   +87.2 &   +49.5 &  60 & st     \\
 30a &  33 &   +25.0 & --126.7 &  65 & st     & &
 79b &  83 & --164.6 &   +43.9 &  50 & st     & &
118b & 123 &   +88.5 &   +51.4 &  90 &        \\
 30b &  33 &   +22.9 & --124.6 &  80 & st     & &
 80a &  84 & --156.6 &   +25.8 &  80 & st     & &
118c & 123 &   +88.2 &   +55.4 &  70 &        \\
 31  &  34 &   +17.0 & --133.1 &  45 & st     & &
 80b &  84 & --158.4 &   +27.4 &  75 & st     & &
118d & 123 &   +85.0 &   +56.7 &  95 & st     \\
 32  &  35 &   +19.7 & --153.1 &  45 & st     & &
 81  &  85 & --157.1 &   +66.9 &  70 &        & &
119a & 124 &   +90.1 &   +48.7 &  65 & st     \\
 33  &  36 &    +6.9 & --160.3 &  55 & st     & &
 82  &  86 & --151.8 &   +63.1 &  70 &        & &
119b & 124 &   +91.7 &   +46.9 &  80 & st     \\
 34  &  37 &    +8.0 & --165.7 &  55 &        & &
 83  &  87 &  --47.0 &   +73.0 &  55 & st     & &
120a & 125 &  +104.2 &   +41.0 &  60 & st     \\
 35a &  38 &   --5.1 & --176.1 & 100 &        & &  
 84  &  88 &  --94.2 &  +121.8 &  45 & st     & &
120b & 125 &  +105.3 &   +45.5 &  80 & st     \\
 35b &  38 &   --7.8 & --179.3 & 100 &        & &
 85  &  89 &  --28.3 &   +73.5 &  95 &        & &
120c & 125 &  +104.2 &   +47.7 &  80 & st     \\
 36  &  39 &    +3.7 & --236.6 &  85 & st     & &
 86  &  90 &    +1.6 &   +81.0 &  60 &        & &
120d & 125 &  +102.4 &   +47.4 &  55 & st     \\
 37  &  40 &    +5.0 & --235.0 &  70 & st     & &
 87  &  91 &  --23.2 &   +82.9 &  70 & st     & &
121a & 126 &  +113.3 &   +44.5 & 100 &        \\
 38  &  41 &    +6.1 & --231.5 & 110 & st     & &
 88  &  92 &  --12.0 &   +87.7 & 130 &        & &
121b & 126 &  +116.8 &   +43.7 & 100 &        \\
 39  &  42 &  --52.3 & --178.5 &  65 & st     & &
 89  &  93 &   --8.0 &   +97.5 &  80 & st     & &
122  & 127 &  +125.8 &   +40.7 &  65 & st     \\
 40a &  43 &  --30.7 & --201.9 &  65 & st     & &
 90  &  94 &   +12.5 &  +129.0 &  65 & st     & &
123a & 128 &  +134.1 &   +38.6 &  65 & st     \\
 40b &  43 &  --33.1 & --202.7 &  60 & st     & &
 91  &  95 &   --1.4 &  +158.9 &  50 & st     & &
123b & 128 &  +133.3 &   +42.1 & 100 & st     \\
 41  &  44 &  --41.9 & --201.9 &  75 & st     & &
 92  &  97 &   --6.2 &  +200.7 &  50 & st     & &
123c & 128 &  +130.4 &   +41.5 &  80 &        \\
 42a &  45 &  --38.4 & --253.9 &  60 & st     & &
 93  &  98 &    +1.3 &  +203.4 &  45 & st     & &
124a & 129 &  +118.1 &   +33.3 &  75 & st     \\
 42b &  45 &  --39.0 & --252.6 &  70 & st     & &
 94a &  99 &    +1.0 &   +57.0 &  55 & st     & &
124b & 129 &  +117.8 &   +35.9 &  60 & st     \\
 43  &  46 &  --43.8 & --256.6 &  75 & st     & &
 94b &  99 &    +2.1 &   +61.0 &  95 &        & &
124c & 129 &  +116.0 &   +38.3 &  90 & st     \\
 44  &  47 &  --45.1 & --261.9 &  35 & st     & &
 95a & 100 &    +1.0 &   +51.1 &  70 & st     & &
125a & 130 &  +114.6 &   +28.2 &  60 & st     \\
 45  &  48 &  --95.0 & --255.0 &  55 & st     & &
 95b & 100 &    +3.2 &   +49.3 &  60 & st     & &
125b & 130 &  +115.7 &   +29.8 &  65 & st     \\
 46  &  49 & --102.2 & --263.8 &  45 & st     & &
 95c & 100 &    +4.8 &   +50.1 &  70 & st     & &
126  & 131 &  +123.7 &   +27.4 &  65 & st     \\
 47  &  50 &  --21.1 &  --41.9 &  60 & st     & &
 96  & 101 &    +9.6 &   +51.1 &  70 & st     & &
127  & 132 &  +124.2 &   +19.7 &  65 & st     \\
\hline
\end{tabular}}
\end{center}
\begin{flushleft}
$^a$ ID number by \citet{belley1992}. $^b$ Offsets from the galactic 
centre, positive to the north and west. $^c$ Star-like profile.
\end{flushleft}
\end{table*}

\begin{table*}
\caption[]{\label{table:tree}
Hierarchical structures of star formation regions.
}
\begin{center}
\scriptsize{
\begin{tabular}{ccccccccccccccc} \hline \hline
Level & Level & Level & Level & & 
Level & Level & Level & Level & Level & & 
Level & Level & Level & Level \\
  1   &   2   &   3   &   4   &       & 
  1   &   2   &   3   &   4   &   5   &      & 
  1   &   2   &   3   &   4   \\
\cline{1-4}\cline{6-10}\cline{12-15}
1a-d   & 1 (230)$^a$    & 1,2 (605)  &  &  & 
44     &                &       &       & & & 
86     & 86h (160)      &       & \\
2      & 2h$^b$ (210)   &       &       &  & 
45     &                &       &       & & & 
87     & 87h (230)      &       & \\
3      &                &       &       & & 
46     & 46h (170)      &       &       & & & 
88     & 88h (230)      & 88,89 (430) & \\
4      & 4,7 (250)      & 4-7 (600) &        & & 
47     &                & 47-49 (700) &       & & & 
89     &                &       & \\
5      & 5,6 (385)      &       &       & & 
48     & 48h (290)      &       &       & & & 
90     &                &       & \\
6      &                &       &       & & 
49     & 49h (140)      &       &       & & & 
91     &                &       & \\
7      &                &       &       & & 
50     & 50h (605)      & 50,51 (995) &       & & & 
92     & 92,93 (365)    &       & \\
8      &                &       &       & & 
51     & 51h (195)      &       &       & & & 
93     &                &       & \\
9a,b   & 9 (110)        & 9h (300) &        & & 
52a-c  & 52,53 (340)    & 52-56 (605) &       & & & 
94a,b  & 94 (175)       & 94-96 (605) & \\
10     & 10h (230)      &       &       & & 
53a,b  &                &       &       & & & 
95a-c  & 95 (240)       &       & \\
11a,b  & 11 (100)       & 11h (235) &        & & 
54a,b  & 54 (130)       &       &       & & & 
96     &                &       & \\
12     & 12h (195)      &       &       & & 
55     & 55,56 (260)    &       &       & & & 
97     & 97h (165)      & 97-99 (525) & 97-99, \\
13a,b  & 13 (75)        &       &       & & 
56     &                &       &       & & & 
98     &                &       & 109 (800) \\
14     & 14,15 (155)    &       &       & & 
57     & 57h (485)      &       &       & & & 
99     &                &       & \\
15     &                &       &       & & 
58     & 58h (395)      &       &       & & & 
100a,b &                & 100-102 (600) & \\
16     & 16h (155)      & 16-18 (620) &        & & 
59a,b  & 59 (130)       &       &       & & & 
101a,b & 101,102 (295)  &       & \\
17     & 17h (185)      &       &       & & 
60a,b  & 60 (115)       &       &       & & & 
102a-c &                &       & \\
18     & 18h (165)      &       &       & & 
61     &                &       &       & & & 
103    & 103h (295)     &       & \\
19     & 19h (280)      &       &       & & 
62a,b  & 62 (100)       &       &       & & & 
104a,b & 104 (265)      &       & \\
20a,b  & 20 (200)       &       &       & & 
63     & 63h (290)      & 63,64 (495) &       & & & 
105a-c & 105 (195)      &       & \\
21     &                &       &       & & 
64     &                &       &       & & & 
106a,b & 106 (135)      & 106h (415) & \\
22     & 22h (280)      &       &       & & 
65a,b  & 65,66 (515)    &       &       & & & 
107    & 107h (335)     &       & \\
23     &                &       &       & & 
66     &                &       &       & & & 
108    &                &       & \\
24     & 24h (300)      &       &       & & 
67     & 67h (230)      &       &       & & & 
109a-c & 109 (290)      &       & \\
25a-c  & 25 (180)       &       &       & & 
68     & 68,69 (360)    &       &       & & & 
110a,b & 110 (205)      &       & \\
26a-c  & 26 (175)       &       &       & & 
69     &                &       &       & & & 
111    &                &       & \\
27     & 27h (440)      &       &       & & 
70     & 70h (305)      &       &       & & & 
112    &                & 112-115 (610) & \\
28     &                &       &       & & 
71     &                &       &       & & & 
113a-d & 113 (225)      &       & \\
29     & 29h (245)      & 29,30 (360) & 29-31 & & 
72     &                & 72-74 (930) &       & & & 
114    &                &       & \\
30a,b  & 30 (140)       &       & (480) & & 
73     & 73h (270)      &       &       & & & 
115a,b & 115 (80)       &       & \\
31     &                &       &        & & 
74     &                &       &       & & & 
116a-d & 116 (400)      & 116,117 (590) & \\
32     &                &       &       & & 
75     & 75h (270)      & 75,76 (670) & & 75-80 & & 
117    & 117h (340)     &       & \\
33     & 33,34 (245)    &       & 33-35 & & 
76     & 76h (320)      &       &       & (2150) & & 
118a-d & 118,119 (445)  &       & \\
34     &                &       & (470) & & 
77     & 77h (215)      &       &       & & & 
119a,b &                &       & \\
35a,b  & 35 (220)       & 35h (270) &        & & 
78     &                &       & 78-80 & & & 
120a-d & 120 (255)      & 120-127 (1145) & \\
36     & 36-38 (270)    & 36-38h (655) &       & & 
79a,b  & 79 (265)       & 79h (545) & (875) & & & 
121a,b & 121 (225)      &       & \\
37     &                &       &       & & 
80a,b  & 80 (120)       &       &       & & & 
122    & 122,123 (340)  &       & \\
38     &                &       &       & & 
81     & 81,82 (335)    &       &       & & & 
123a-c &                &       & \\
39     & 39h (165)      &       &       & & 
82     &                &       &       & & & 
124a-c & 124,125 (450)  &       & \\
40a,b  & 40 (115)       & 40,41 (490) &        & & 
83     &                &       &       & & & 
125a,b &                &       & \\
41     & 41h (140)      &       &       & & 
84     &                &       &       & & & 
126    &                &       & \\
42a,b  & 42-44 (190)    & 42-44h (585) &        & & 
85     & 85h (315)      &       & 85,87-89 & & & 
127    & 127h (155)     &       & \\
43     &                &       &       & & 
       &                &       & (695) & & & 
       &                &       & \\
\hline
\end{tabular}}
\end{center}
\begin{flushleft}
$^a$ Diameter. $^b$ Star formation region with halo.
\end{flushleft}
\end{table*}

\end{document}